\documentclass[twocolumn,twoside,slac]{revtex4}
\usepackage{graphicx}
\usepackage{fancyhdr}
\begin{document}

\title{String Parton Models in Geant4}

%

\author{G.Folger, J.P.Wellisch}
\affiliation{CERN, CH-1211 Geneva, Switzerland}

\begin{abstract}
Dual parton or quark gluon string model are the by now almost
standard theoretical techniques by which one can arrive at precision
description of high energy, soft, inclusive reactions. These reactions make the
part of jets at energies that contribute strongly to discovery channels such as
H$\rightarrow$WWjj, or search for compositeness at the highest transverse momenta. The
above modeling approach is available with Geant4 for nucleon induced reactions
since the first release. Its object oriented design and parameter set was
recently extended to allow for simulation of pion and kaon induced reactions,
as well as heavy ion reactions. We will briefly describe the theory and
algorithmic approaches that underly the modeling, show the object oriented
designs and component structure of the string parton sub-systems of Geant4,
present validation/verification results pertaining to these models, as well as
results concerning their usage in calorimeter simulation.   
\end{abstract}

\maketitle

\thispagestyle{fancy}



\section{Overview}


The string parton models in Geant4 \cite{geant4} serve to simulate inelastic reactions of
high energy particles with nuclei. The Geant4 string parton models are modular.
To simulate the interaction of high energy particle with the nucleus several
building parts are used together, and for some of the parts there is more than
one choice. 

In a first stage the interaction of a high energy particles with at least one
nucleon of the nucleus is modeled using a string excitation model. At the
moment Geant4 provides two different string excitation models, the diffractive
string excitation and the quark gluon string model. In the initial state, a
nucleus is built consisting of individual protons and neutrons, the nucleons.
The result of an interaction between the primary and the nucleus are one or
several excited strings and a nucleus in an excited state. A string consists of
two endpoints with defined quark content and carries energy and momentum.  The
fragmentation of the excited strings into hadrons is handled by a longitudinal
string fragmentation model. The interaction of secondaries with the excited
nucleus will be handled by a cascade model. Until this implementation will be
completed, secondaries are assumed to be produced outside of the nucleus. The
de-excitation of the excited nucleus is further simulated by nuclear
fragmentation, precompound, and nuclear de-excitation models.  

\section{Object Oriented Design Overview}

The string parton model is part of the Geant4 simulation toolkit. The design of
the string parton model was made with a set goals. The use of several models or
implementations is made possible through  the use of common interfaces,
allowing also for easy integration of more models. Common parts between
multiple implementations or between various models are shared in common
classes. As an example the model of the nucleus is shared between many of the
theory driven models for hadronic reactions in Geant4.

\begin{figure*}
\rotatebox{-90}{\resizebox{!}{17cm}{
    \includegraphics{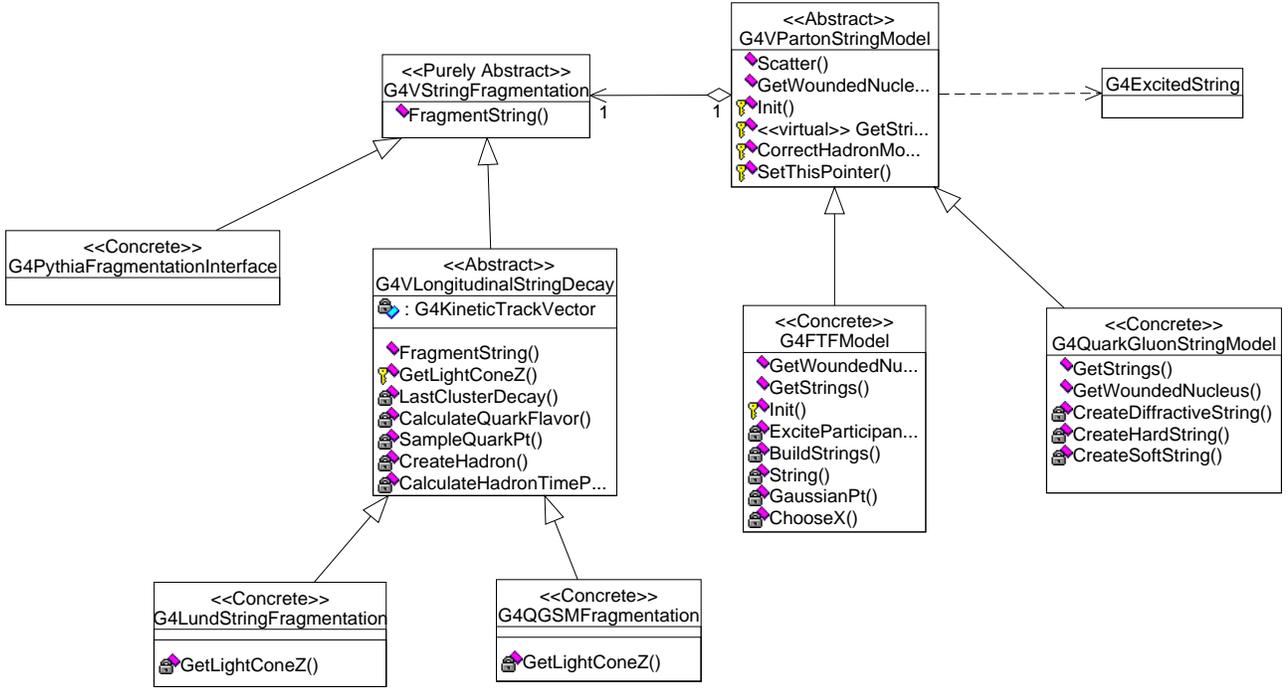}
  }
}
\caption{Overview of design for parton string model classes.}
\label{f:design}
\end{figure*}

An overview of the design of the parton string models of Geant4 using UML
notation in shown in figure \ref{f:design}. The interface of the parton string
models to the Geant4 toolkit is defined and partly implemented in the abstract
{\em G4VPartonStringModel} class. The classes implementing the diffractive
string excitation {\em G4FTFModel} and the quark gluon string model {\em
G4QGSModel} are concrete implementations of the abstract {\em
G4VPartonStringModel} class.  The interface to string fragmentation again is
defined in an abstract class, {\em G4VStringFragmentation}. The model for the
longitudinal string fragmentation is implemented by the {\em
G4VLongitudinalStringDecay} class. The latter is abstract, as the fragmentation
function is not implemented. This shares the algorithm between  concrete
implementations and allows specific string models to use specific fragmentation
functions.  The concrete classes {\em G4LundStringfragmentation} and {\em
G4QGSMFragmentation} are used, respectively,  by the diffractive parton string
and by the quark gluon string models.  Other string fragmentations schemes are
possible, as e.g. indicated by the example of {\em
G4PhythiaFragmentationInterface}. This is foreseen as an interface to the
string fragmentation of Phythia7~\cite{phythia7}

\section{Modeling the Nucleus}

The nucleus is modeled as an ensemble of protons and neutrons.
Each nucleon is positioned randomly in configuration and momentum space.
The positions are chosen at random following to the nuclear density distribution. For heavy
nuclei, ie. nuclei with a mass number above 16, we use a density distribution of
the Wood-Saxon form:
\begin{equation}\label{eq:d_fermi}
\rho (r_i) = \frac{\rho_0} {1 + \exp (\frac{(r_i-R)}{a})}
\end{equation}   
where $R$ and $a$ depend on the mass number of the nucleus.

For light nuclei we use a density distribution from the harmonic oscillator
model:
\begin{equation}\label{eq:d_harmonic}
\rho (r_i) = (\pi R'^{2})^{-3/2} \exp (-r_i^2 / R'^{2}),
\end{equation} 
where $R'$ is the effective nuclear radius, and depends on the mass number of the nucleus.

The sampling in configuration space is done such that no two nucleons have a distance from each other less than $0.8fm$.

The momentum of each nucleon is chosen random in momentum space with a maximum
momentum $p_F^{max}$ 
\begin{equation}
p_F^{max} = \hbar c (3\pi^2 \rho(r_i))^{1/3} 
\end{equation}
which is a function of the density $\rho(r_i)$ obtained from equation \ref{eq:d_fermi} or
\ref{eq:d_harmonic}. Momentum balance is achieved by choosing the direction of
momentum for the nucleons such that the sum of all nucleon momenta is
zero.

For the purpose of further calculations, this nucleus is then collapsed into
two dimensions perpendicular to the direction of the primary particle. This way
we take into account that at high energies the coherence length of the string
fragmentation is large in comparison to the thickness of the  (relativistically
contracted) nucleus. All scattering is hence assumed to happen independent of
any time ordering, and to be correlated only through energy and baryon-number
conservation.

\section{Diffractive scattering model}

The diffractive scattering model simulates the interaction of an high energetic
hadron with a nucleus, where the incident
particle may interact with one or several nucleons in the nucleons. For each
nucleon the impact parameter is calculated, and using the impact parameter and the
interaction center of mass energy, the interaction probability is calculated
from the inelastic and diffractive cross section respectively using the eiconal model. The interacting
nucleons are selected using uniform sampling of the interaction probability. 

The diffractive scattering of the primary particles with a nucleon is modeled using 
an approach similar to the one employed in Fritiof\cite{FRITIOF}. In this approach 
the scattering particles only exchange
momentum:
\begin{equation}
\begin{array}{c}
p'_1 = p_1 + q  \\
p'_2 = p_2 - q
\end{array}
\end{equation}
where $p_{1,2}$ are the momenta of the incoming, and $p'_{1,2}$ the momenta of the
scattered particles, and $q$ is the momentum exchanged. A string is formed
for each of the two scattered particles, using the quark content of the original
hadron by assigning the quarks of the hadron randomly to the two string ends.

In the center of mass system and using  
light-cone coordinates,
the momenta of the incoming particles are 
\begin{equation} 
\begin{array}{c}
p_1 = (E_1^+, m_1^2/E_1^+, {\bf 0}) \\  
p_2 = (E_2^-, m_2^2/E_1^-, {\bf 0})
\end{array}
\end{equation}
and the momentum transfer is 
\begin{equation}
q= ( -q_t^2/x^-E_2^-, q_t^2/x^+E_2^+, {\bf q_t})
\end{equation}

The model does not naturally contain transverse momentum, hence
the transverse momentum ${\bf q_t}$ is sampled from a gaussian distribution
with a default width of $0.8$~GeV, using a simple multiple small-angle scattering assumption.

The longitudinal components $q^+$ and $q^-$ of the momentum exchange are
obtained sampling $x^+$ and $x^-$ from the parton distribution:
\begin{equation}
u(x) = x^\alpha (1-x)^\beta
\end{equation}
where for the diffractive string model the parameters are $\alpha=-1$, and
$\beta=0$.

The masses of the resulting strings must fulfill the kinematic constraint
\begin{equation}
p^{\prime +}_{1,2} \, p^{\prime -}_{1,2} \ \geq \ p_{1,2}^2 + q_t^2
\end{equation}
where $p_{1,2}^2$ are the masses of the incident particle and the nucleon.

\begin{figure}
  \resizebox{!}{7cm}{
    \includegraphics{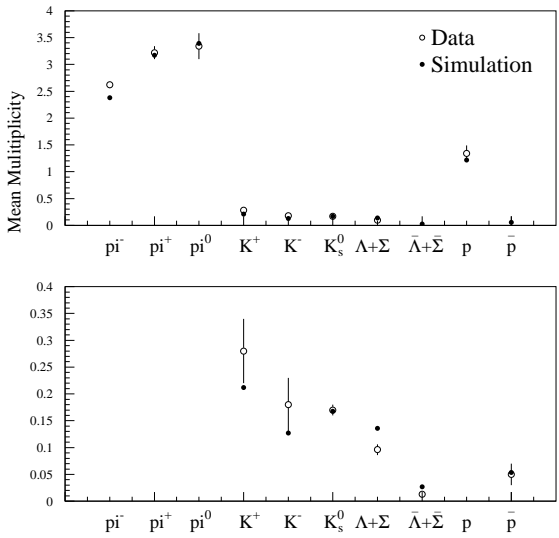}
  }
\caption{Mean multiplicity for reactions $p \ H \rightarrow \ X$ at
$200Gev/c$. Open circles are data, and full circles is Monte Carlo simulation using the
diffractive string model.}
\label{f:ftf}
\end{figure}

As an illustration of result obtained from this model, we plot in figure
\ref{f:ftf} the mean multiplicity for several particles types observed in the
final state in reactions $p \ H \rightarrow X$ at a momentum of $200 GeV/c$ in
laboratory frame for the incoming proton. 

\section{Quark Gluon String Model}

The Quark Gluon string model, too, allows to simulate reactions of high energy hadrons
with nuclei and also to simulate high energy electro- and gamma-nuclear reactions.
Unlike the diffractive models, in this case, the colour flow is assumed to be between partons from the interaction partners. 

The nucleus is modeled as above. The impact parameter for each nucleon $b_{i}$
is calculated, collapsing the nucleus into a plane orthogonal to the incident
primary particle.  The hadron nucleon collision probabilities are calculated
using the cross-sections of the eiconal model and using gaussian
distributions for the wave-functions of both hadrons and nucleons \cite{Capella78}. 
They are used to determine the number of participating nucleons in the nucleus.
In the quark-gluon string model, each hadron-nucleon interaction is assumed to be 
mediated by the exchange
of one or more Pomerons.
Hence for each pair of participants
 the number of Pomerons $n$ is sampled. This is possible as in
the Regge Gribov approach the reaction probability can be factorized,
and the contribution of any pair of participants can be written as a sum 
over the number of Pomerons exchanged:

\begin{equation}
P_{i}(b_{i},s) 
 = \frac{1}{c} (1 - \exp {[-2u(b_{i},s)])} 
 = \sum_{n=1}^{\infty} P_{i}^{(n)}(b{_i},s)
\end{equation}
The individual contribution of the $N$ Pomeron graph here reads as 
\begin{equation}
P_{i}^{(n)}(b{_i},s) = \frac{1}{c} \exp{[ -2u(b_{i}^2,s)] }\frac{(2u(b_{i}^2,s))^n}{n!}
\end{equation}

where the Eikonal can be written as 
\begin{equation}
u(b_{i}^2,s) = \frac{z(s)}{2} \exp(\frac{b_{i}^2}{4\lambda(s)}).
\end{equation}
Here $s$ is the c.m.s. energy; $c$, $z(s)$, and $\lambda(s)$ are
functions of the eiconal model cross section description, that can be expressed
in terms of the Pomeron vertex and trajectory parameters. 

In this model a small fraction of interactions is diffractive, its
probability is split off using Baker's shower enhancement factor $c$ \cite{baker76}:
\begin{equation}
P_{i}^{diff}(b_{i},s) = \frac{1-c}{c}(P_{i}^{tot}(b_{i},s) - P_{i}(b_{i},s))
\end{equation}

Strings are formed using the parton exchange mechanism by sampling of parton
densities and ordering pairs of partons into color coupled entities
\cite{Kaidalov82}.
Each Pomeron is treated as a pair of colour triplet strings, where the string ends are
attached to partons in the interacting hadrons. Strings are then decayed as described later
in this paper. The relative contributions from valence and sea quarks are split, so that the
fragmentation functions will look like
\begin{eqnarray*}
\phi_n^h=a^h[F_v^h(x_+,n)F_{anti-v}^h(x_-,n)\\
+F_v^h(x_-,n)F_{anti-v}^h(x_+,n)\\
+ (n-1)(F_s^h(x_+,n)F_{anti-s}^h(x_-,n)\\
+F_s^h(x_-,n)F_{anti-s}^h(x_+,n))]
\end{eqnarray*}
where $v$ and $s$ stand for valence and see respectively, and 
the functions $F$ are the parton density functions folded with the fragmentation functions and the
transverse momentum function:
\begin{equation}
F^h(x_\pm,n)=\sum_i\int_{x_\pm}^1f_i(x',n)G_i^h(x_\pm/x')T(p_T,n) dx'
\end{equation}

\begin{figure}
  \resizebox{!}{8.2cm}{
    \includegraphics{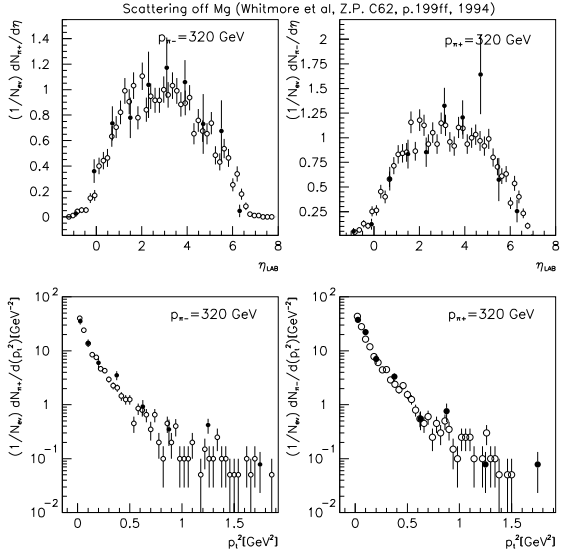}
  }
\caption{
Comparison of data and Monte Carlo prediction for the quark gluon string model.
We show rapidity and transverse momentum square distributions of $\pi^+$ produced
in pion Magnesium reactions at 320 GeV. Open circles are the Monte Carlo predictions, and points are
experimental data.
}
\label{qgs1}
\end{figure}

\begin{figure}
  \resizebox{!}{6cm}{
    \includegraphics{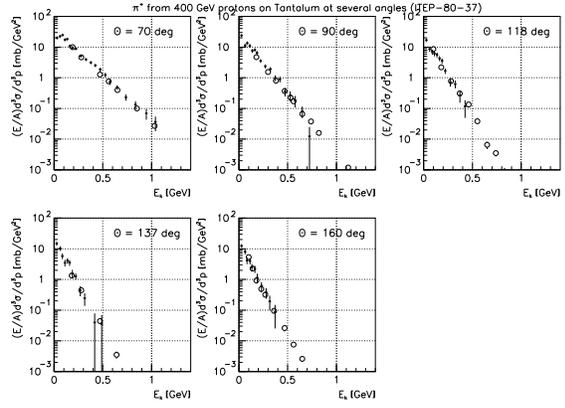}
  }
\caption{
Comparison of data and Monte Carlo prediction for the quark gluon string model.
We show invariant cross-sections of $\pi^+$ produced
in pion Magnesium reactions at 400 GeV as a function of the pion kinetic energy. 
Each plot describes a different scattering angle.
Open circles are experimental data, and points are Monte Carlo prediction.
}
\label{qgs2}
\end{figure}

Examples of the predictive power of this model are given in Fig.\ref{qgs1} and \ref{qgs2}.

\subsection{Electro- and gamma-nuclear interactions}

The quark gluon string model is also used to simulate electro- and gamma nuclear reactions.
This is done using a single interaction assumption and vector meson dominance. From there,
the quark-gluon string model can be applied as for any other hadron, once the Pomeron vertex and slope 
parameters are properly adjusted.
For electro-nuclear reactions, in addition, it is of course necessary to assume an equivalent photon flux
mediating the reactions, and to take into account the virtuality of the photons.
The cross section for a high energy gamma ray to interact with a nucleus, and all aspects of the
equivalent photon flux hypothesis and momentum transfer calculations is
presented in \cite{hpw-1}. 

\section{String fragmentation}

The string as created by the diffractive or the quark gluon string model is
characterized by its four momentum and its constituents, i.e. the quark
contents at the two endpoints of the string. The algorithm for hadronisation of
the string is common for both string models, except for the fragmentation
function used. The string repeatedly is split into a hadron and a new string,
until the energy in the strings gets too low for further splitting.

In the current implementation, a constituent can be a up, down, or strange
quark or antiquark, or a diquark or anti-diquark of up, down, or strange
quarks. The strings must have integer charge, so only the following combinations
of constituents plus the charge conjugated combinations are allowed:
$q-\overline{q}$, $q-(qq)$, $(qq)-\overline{(qq)}$.  In the longitudinal
fragmentation model the constituents move in opposite direction increasing the
tension on the string. The string then breaks creating a new quark - antiquark
$q-\overline{q}$ or diquark - anti-diquark $(qq)-\overline{(qq)}$ pair. The
different quark flavours are created with a relative probability of: 
\begin{equation} u \ : \ d \ : \ s \  = \ 1 \ : \  1  \ : \ 0.27   
\end{equation}  
Diquark - anti-diquark pairs are produced in 10\% of all cases. 

Half of the newly created pair forms a hadron with one of the constituents, and the
other half of the newly created pair together with the remaining constituent
forms a new string. 

The quark content gives the charge of the hadron and its type.  For mesons we
create scalar and vector mesons taking into account the mixings of neutral
mesons. For barions we construct barions from the lowest SU(3) octet (spin 1/2
barions) and from the lowest SU(3) decuplet (spin 3/2 barions).     

The quark or diquark get transverse momentum sampled from a gaussian
distribution using a width of $\sigma = 0.5 GeV$. The antiquark or anti-diquark
gets the opposite transverse momentum to conserve total transverse momentum.

The longitudinal momentum is split off the string longitudinal momentum using a fraction
$z$ sampled from a
fragmentation function. For the diffractive string model we use the 
Lund fragmentation function
\begin{equation}
f(z,m_H,q_t) = \frac{1-z}{z} \exp( \frac{-b (m_H^2 + q_t^2)} {z} )
\end{equation}
where $m_H$ is the mass of the created hadron, and $q_t$ is the transverse
momentum of the hadron. For the quark gluon string model, we use \cite{Kaidalov87}
\begin{equation}
f^h(z,q_t) = [ 1 + \alpha^h(<q_t>)] (1-z)^{\alpha^h(<q_t>)}
\end{equation}
where the parameter $\alpha$ depends on the type of newly created quark or diquark, and the average transverse momentum.
Finally, the transverse momentum of the hadron is the sum of
transverse momenta of the string constituent and of the newly created quark.
  
This process is iterated until the energy of the string
gets too low to form further hadrons.

\begin{acknowledgments}
The authors wish to thank CERN for their support and A.B.Kaidalov for few but very useful discussions.

\end{acknowledgments}


\end{document}